\documentstyle[11pt]{article}
\def\ra{\rightarrow}

\let\a=\alpha
\let\b=\beta 
\let\c=\gamma
\let\C=\Gamma  
\let\d=\delta
\let\e=\epsilon
\def\eb{{\bar\epsilon}}
\def\tb{{\bar\theta}}
\def\kb{{\bar\kappa}}

\def\ce{{\cal\epsilon}}
\def\ve{\varepsilon}

\def\tr{{\rm tr}}
\def\det{{\rm det}}

\let\k=\kappa
\let\l=\lambda
\let\L=\Lambda
\let\m=\mu
\let\n=\nu 
\let\r=\rho
\let\s=\sigma
\let\t=\theta
\let\w=\omega  

\def\bd{\begin{document}} 
\def\ed{\end{document}}
\def\be{\begin{equation}}
\def\ee{\end{equation}}
\def\ba{\begin{array}}
\def\ea{\end{array}}
\def\bea{\begin{eqnarray}}
\def\eea{\end{eqnarray}}
\def\nn{\nonumber}
\def\ni{\noindent} 
\let\la=\label 
\let\bm=\bibitem
\def\p{\partial}
\def\dt{d\tau_1\cdots d\tau_n}
\def\pn{(\partial_1+\cdots + \partial_n)}
\def\lra{\leftrightarrow}
\def\ra{\rightarrow}
\def\qq{\quad\quad}
\def\mod{{\rm mod}}

\def\ft#1#2{{\textstyle{{\scriptstyle #1}
\over {\scriptstyle #2}}}}
\def\fft#1#2{{#1 \over #2}}
\def\sst#1{{\scriptscriptstyle #1}}
\newcommand{\eq}[1]{(\ref{#1})}
\def\eqs#1#2{(\ref{#1}-\ref{#2})}
\def\cites#1#2{\cite{#1}-\cite{#2}}
\def\Hat#1{\widehat{#1}}
\def\nosum{({\rm no\ sum\ over}\ i~)}

\def\pl#1#2#3{Phys.~Lett.~{\bf {#1}B} (19{#2}) #3}
\def\np#1#2#3{Nucl.~Phys.~{\bf B{#1}} (19{#2}) #3}
\def\cqg#1#2#3{Class.~and Quant.~Gr.~{\bf {#1}} (19{#2}) #3}
\def\mpl#1#2#3{Mod.~Phys.~Lett.~{\bf A{#1}} (19{#2}) #3}
\def\pr#1#2#3{Phys.~Rev.~{\bf {#1}D} (19{#2}) #3}
\def\prep#1#2#3{Phys.~Rep.~{\bf {#1}C} (19{#2}) #3}

\begin{document}

\begin{titlepage}

\hfill{CTP TAMU-45/97}

\hfill{hep-th/9711128}

\vspace{30pt}


\centerline {\Large\bf Superparticles, p-Form Coordinates }
\medskip
\centerline{\Large\bf and the BPS Condition}


\vspace{30pt}

\centerline{{\large I. Rudychev and E. Sezgin}\footnote{Research
supported in part by NSF Grant PHY-9722090. }}

\vspace{15pt}

\centerline{\it Center for Theoretical Physics, Texas A\&M University,}
\centerline{\it College Station, Texas 77843, U.S.A.}

\vspace{50pt}


\centerline{ABSTRACT}


\vspace{15pt}

A model for $n$ superparticles in $(d-n,n)$ dimensions is studied. The
target space supersymmetry involves a product of $n$ momentum
generators, and the action has $n(n+1)/2$ local bosonic symmetries and
$n$ local fermionic symmetries. The precise relation between the
symmetries presented here and those existing in the literature is
explained. A new model is proposed for superparticles in arbitrary
dimensions where coordinates are associated with all the $p$-form
charges occuring in the superalgebra. The model naturally gives rise to
the BPS condition for the charges.

\end{titlepage}


\section{ Introduction }


Consider supercharges $Q_\a$ in $(d-n,n)$ dimensions where $n$ is the
number of time-like directions. Let the index $\a$ label the minimum
dimensional spinor of $SO(d-n,n)$. In the case of extended
supersymmetry, the index labelling the fundamental representation of the
automorphism group is to be included. We will suppress that label for
simplicity in notation. Quite generally, we can contemplate the
superalgebra
\bea
\{ Q_\a, Q_\b \} &=& Z_{\a\b}\ , \nn\\
 \left[Z_{\a\b}, Z_{\c\d}\right] &=& 0 \ , \la{alg}
\eea
where $Z_{\a\b}$ is a symmetric matrix which can be expanded in terms
of suitable $p$-form generators in any dimension by consulting Table 3
provided in Appendix B. The $Z$ generators have the obvious commutator
with the Lorentz generators, which are understood to be a part of the
superalgebra.

Next, consider dimensions in which the $n$\,th rank $\c$-matrix is
symmetric
\footnote
{For example, we can consider
$(n,n)$ and $(8+n,n)$ dimensions where one can have
(pseudo) Majorana-Weyl spinors, or $(4+n,n)$ dimensions where (pseudo)
symplectic Majorana-Weyl spinors are possible. In the latter case, the
tensor product of the antisymmetric $n$'th rank $\c$-matrices with the
antisymmetric invariant tensor of the symplectic automorphism group is
symmetric. The spinors need not be Weyl, and thus we can consider other
dimensions as well (see Appendix B). 
}.
In $(d-n,n)$ dimensions, setting the $n$\,th rank generator equal to a
product of $n$ momentum generators \cite{b2} and all the other $Z$
generators equal to zero, one has
\be 
\{ Q_\a, Q_\b \} = 
	(\c^{\m_1\cdots \m_n})_{\a\b}~P_{\m_1}^1 \cdots P_{\m_n}^n 
\la{algn}\ .
\ee
Motivations for considering these algebras have been discussed elsewhere
(see \cite{b2,bk2,rs1,rss}, for example). The fact they can be realized
in terms of multi-particle systems was pointed out in \cite{bk2}.
Superparticle models were constructed in \cite{rs1} and \cite{deli1},
with emphasis on the cases of $n=2,3$. In \cite{rs1} a multi-time model
was considered in which a particle of type $i=1,2,3$ depended on time
$\tau_i$ only. As a special case, an action for a superaparticle in the
background of one or two other superparticles with constant momenta was
obtained. In the model of \cite{deli1}, where single time dependence was
introduced, a similar system was described. The results agree, after one
takes into account the trivial symmetry transformations that depend on
the equations of motion.

These models were improved significantly in \cite{deli2}, where all
particles are taken to have arbitrary (single) time dependence. In
\cite{deli2}, it was observed that the theory had $n$ first class
fermionic constraints, but one fermionic symmetry was exhibited (or $m$
of them for extended supersymmetry with $N=1,...,m$). Furthermore, the
form of the transformation rules given in \cite{deli2} appear to be
rather different than those found earlier in \cite{rs1} (albeit in the
context of a restricted version of the model).

The purpose of this note is three-folds: (1) to provide the full
$\k$-symmetry transformations of the action considered in \cite{deli2},
(2) to show the precise relation between the single (or $m$)
$\k$-symmetry parameter of \cite{deli2} and the $n$ (or $m\times n$)
such parameters here, as well as the relation between the bosonic and
fermionic transformations of \cite{deli2} and those given here and (3)
to present a new model for superparticle in which coordinates are
associated with the supercharges and {\it all} the bosonic $p$-form
generators occuring in their anticommutator.

The new model provides a realization of the $p$-form charges and it
gives rise to the BPS condition. In this model, we introduce the
spinorial symmetric matrix coordinates $X^{\a\b}$ corresponding to the
generators $Z_{\a\b}$ and the associated momentum variables $P_{\a\b}$
\footnote{
The notation $X^{\a\b}$ has been used in the literature for situations
where it stands for $X^{\a\b}\equiv X^\m \c_\m^{\a\b}$ (see, for
example, \cite{bandos}), whereas in our model $X^{\a\b}$ contains {\it
all} the symmetric $\c$-matrices that can occur in the expansion. A
superparticle model was proposed in \cite{lukierski} in which extra
bosonic coordinates were introduced only for the usual central charges
that are singlets of the Lorentz group. Coordinates for $p$-forms were
used in \cite{tc} in the context of a search for string theory in eleven
dimensions, and in the context of twistor superparticle action in ten
dimensions \cite{sol} (We thank D. Sorokin for pointing this reference
to us). Coordinates for super $p$-form charges in general were
introduced in \cite{bs2,es2}, in the construction of generalized super
$p$-brane actions. The new superparticle model presented here does not
correspond to the particle limit of these brane actions. Morover, the
constraints of our model differ from those arising in the twistor
supercparticle model of \cite{sol}. The $p$-form coordinates have also
arisen in the context of a free differential algebra \cite{df} extension
of the usual supertranslation group in which the $p$-form potentials
that are required in the construction may be viewed as coordinates on an
extended group manifold \cite{t1}.
}.
In addition, we introduce the Lagrange multiplier variables $e_{\a\b}$,
which are symmetric or antisymmetric, depending on the signature of
spacetime. The model exhibits a reach set of local and global symmetries
and the Lagrange multiplier equation of motion gives rise to a general
on-shell condition for the momenta, which contains the BPS condition. An
example of this model is given for $(2,2)$ dimensional target space, and
it contains the coordinates $X^\m$ and $X^{\m\n}$.

The new model is rather universal and it works in arbitrary dimensions.
We will see that there exists a particular reduction of the model which
resembles the multi-superparticle model discussed in Sec. 2. The new
model is hoped to provide a general scheme for realizing supersymmetry
in dimensions beyond eleven, and to be more suitable for a brany
generalization, in comparison with the previously considered
superparticle models.

\bigskip
 

\section{ The Model for $n$ Superparticles in $(d-n,n)$ Dimensions}


Consider $n$ superparticles which propagate in $(d-n,n)$ dimensional
spacetime. Let the superspace coordinates of the particles be denoted by
$X_i^\mu (\tau)$ and $\theta_i^\a(\tau)$\, with $i=1,...,n$,
$\m=0,1,...,d-1$ and $\a = 1,..., dim\, Q$, where $dim\ Q $ is the
minimum real dimension of an $SO(d-n,n)$ or $SO(d-n,n)\times G$ spinor,
with $G$ being the automorphism group. Working in first order formalism,
one also introduces the momentum variables $P^i_\m(\tau)$ and the Lagrange
multipliers $e_{ij}(\tau)$.

The superalgebra \eq{algn} can be realized in terms of the supercharge 
\be
Q_\a= \p_\a +\C_{\a\b}~\t^\b\ ,   \la{qn} 
\ee 
where, using the notation of \cite{deli2}, we have defined 
\be
\C  \equiv \frac{1}{n!}~\ve_{i_1\cdots i_n}~\c^{\m_1\cdots \m_n}~
	P_{\m_1}^{i_1} \cdots P_{\m_n}^{i_n}\ . \la{defc}
\ee 
The spinorial derivative is defined as $\p_\a =\p/\p\t^\a $, acting
from the right. The transformations generated by the supercharges
$Q_\a$ are \cite{deli2}
\be
\d_\ce X_i^\m = - \eb~V_i^\m \t \ , \qq
\d_\ce \t = \ce\ ,\qq
\d_\ce  P_i^\m =  0\ , \qq
\d_\ce e_{ij} = 0\ ,\la{sn} 
\ee
where, in the notation of \cite{deli2}, we have the definition 
\be 
 V_i^\m \equiv \frac{1}{(n-1)!}~\ve_{i i_2\cdots i_n}~
	\c^{\m \m_2\cdots \m_n}~ P_{\m_2}^{i_2}\cdots P_{\m_n}^{i_n}\ .
\la{defv}
\ee 
We use a convention in which all fermionic bilinears involve $\c
C$-matrices with the charge conjugation matrix $C$ suppressed, e.g. $\tb
\c^{\m\n} d\t \equiv  \t^\a (\c^{\m\n}C)_{\a\b}\,d\t^\b$. Otherwise
(i.e. when there are free fermionic indices), it is understood that
the matrix multiplications involve northeast-southwest contractions,
e.g. $(\C\,\t)_\a \equiv (\C)_\a{}^\b \t_\b$. Note that
\be
P^i_\m V_i^\m = n \C\ ,\quad\quad V_i^\m~dP^i_\m =d\C\ ,\quad\quad 
P^i_\m~dV_i^\m =(n-1) d\C\ ,
 \la{lemma1}
\ee
where $d \equiv \p /\p\tau$.

The action constructed in \cite{deli2} is given by
\be
I = \int d \tau~\left( P^i_\m \Pi_i^\m 
	-\ft12 e_{ij} P^i_\m P^{j\m} \right)\ , \la{ract}
\ee
where
\be
\Pi_i^\m = d X_i^\m +\ft{1}{n}\tb V_i^\m d \t \ . \la{le}
\ee
It should be noted that the line element \eq{le} is not invariant under
supersymmetry, but it is defined such that $P^i_\m \Pi_i^\m$ transforms
into a total derivative as $P^i_\m \left(\d_\ce \Pi_i^\m\right)=(1-n)
d(\eb~\C \t)$. Consequently, the action \eq{ract} is invariant under the
global supersymmetry transformations \eq{sn}. It also has the local
bosonic symmetry
\be
\d_\L e_{ij} =  d \L_{ij}\ , \qq 
\d_\L X_i^\m = \L_{ij} P^{j\m} \ , \qq
\d_\L P_i^\m = 0 \ , \qq
\d_\L \t     = 0 \ , \la{bn}
\ee
where the transformation parameters have the time dependence
$\L_{ij}(\tau)$. These transformations are equivalent to those given in
\cite{deli2} by allowing gauge transformations that depend on the
equations of motion, as will be shown in Appendix A. 

The action \eq{ract} is also invariant under the following 
$\k$-symmetry transformations
\bea
\d_\k \t &=& \c^\m \k_i P^i_\m \ ,\nn\\
\d_\k X_i^\m &=& -\tb V_i^\m (\d_\k \t) \ , \nn\\
\d_\k P_i^\m &=& 0\ , \nn\\
\d_\k e_{ij} &=& -\ft{4}{(n+9)}~\kb_{(i}~V_{j)}^\m \c_\m~d\t \ .
\la{rk}
\eea
In showing the $\k$-symmetry, the following lemmas are
useful:
\bea
P^i_\m \left(\d_\k \Pi_i^\m\right) &=&
\ft{(1-n)}{n}~d \left( \tb_k~\C \d_\k \t_k \right)
+ \ft{2}{n} (\d_\k \tb_k)~\C~d \t_k \ ,
\la{lemma2}\\[+0.25cm] 
\C \c^\m P^i_\m &=& \ft{1}{(n+9)}~M^{ij} V_j^\m \c_\m\ ,
\la{lemma3}
\eea
where
\be
M^{ij} \equiv P^i_\m P^{j\m}\ . \la{mij}
\ee

A special combination of the transformations \eq{rk} was found in
\cite{deli2}, where it was also observed that there is a total of $n$
first class constraints in the model. The $\kappa$-symmetry
transfomations given above realize the symmetries generated by these
constraints. 

The commutator of two $\k$-transformations closes on-shell onto the
$\L$-transformations 
\be
[\d_{\k_{(1)}},\d_{\k_{(2)}}]= \d_{\L_{(12)}} \ , \la{close}
\ee
with the composite parameter given by
\be
\L^{(12)}_{ij}= -4\kb^{(2)}_{(i}~\C~\k^{(1)}_{j)}\ . \la{rc}
\ee
It is clear that the remaining part of the algebra is
$~[\d_\k,\d_\L]=0~$ and $~[\d_{\L_1},\d_{\L_2}]=0~$.

Finally, we note that the field equations following from the action
\eq{ract} take the form
\be
P^i_\m P^{j\m} = 0\ ,\qq
dP^i_\m = 0\ , \qq
\C d\t  = 0\ , \qq
dX_i^\m + \tb V_i^\m d\t -e_{ij} P^{j\m}=0 \ . \la{reom}
\ee

The precise relation between the bosonic symmetry transformations
\eq{bn}, the $\kappa$-symmetry transformations \eq{rk}, and those
presented in \cite{deli2} is discussed in Appendix A. 

In this section we have used a notation suitable to simple (i.e. $N=1$)
supersymmetry in the target superspace. One can easily account the
extended supersymmetry case by introducing an extra index $A=1,...,m$
for the fermionic variables, thereby letting $\e \ra \e^A$, $\theta \ra
\theta^A$ and $\k^i\ra \k^{Ai}$, etc. All the formulae of this section
still hold, since no need arises for any Fierz rearrangents that might
potentially put restrictions on the dimensionality of the target
superspace.

\bigskip


\section{A New Model and the BPS Condition}

\subsection{A Universal Model}


The model described above picks out the $n$\,th rank antisymmetric
bosonic generator in the algebra \eq{alg}. Furthermore, a polynomial in
momenta occurs in the definition of the `torsion' tensor, making it
difficult to interpret the model in curved space. This led us to
consider a different realization of the algebra \eq{alg}. We take the
most democratic point of view and define a generalized superspace in
which coordinates are associated with $Q_\a$ and $Z_{\a\b}$:
$$
(Q_\a\ ,\  Z_{\a\b}) \ \ \ra \ \ (\t^\a\ , \ X^{\a\b} ) \ . \la{ma}
$$
In addition, we introduce the momenta $P_{\a\b}$ (symmetric) and
the Lagrange multipliers $e_{\a\b}$ with the same symmetry property of
the charge conjugation matrix (see Appendix B):
\be
e_{\a\b}=\e_0\,e_{\b\a}\ ,\qq   C^T=\e_0\,C\ . \la{ecs}
\ee
The enlargement of the superspace at this scale, where the bosonic
coordinates do not necessarily include the familiar Lorentz vector
$X^\m$ need not alarm us, because the model may allow reductions to
familiar settings in lower dimensions, and at the same time give rise to
new physical situations. 

Having defined the basic fields of the model, we propose the following
action
\be
I= \int d\tau \left( P_{\a\b}\, \Pi^{\a\b} 
+\ft12 e_{\a\b}\, (P^2)^{\a\b} \right) \ ,
\la{na}
\ee
where
\be
\Pi^{\a\b} = dX^{\a\b} - \t^{(\a} d\t^{\b)} \ ,
\ee
and $(P^2)^{\a\b}\equiv P^{\a\c} P_\c{}^\b$. The raising and lowering of
spinor indices is with charge conjugation matrix $C$ which has 
symmetry property \eq{ecs}.

The bosonic symmetry of the action takes the form
\be
\d_\L e_{\a\b} = d\L_{\a\b}\ , \qq
\d_\L X^{\a\b} = -\L^{(\a}{}_\c\, P^{\b)\c}\ ,   \qq
\d_\L P_{\a\b} = 0 \ , \qq
\d_\L \t = 0 \ .  \la{nl}
\ee 

In addition, there is a trivial bosonic $\Sigma$-symmetry given by
\be
\d_\Sigma e_{\a\b} = \ft12 \left( \Sigma_{\c\a}~P_\b{}^\c
	+\e_0\,\Sigma_{\c\b}~P_\a{}^\c\right)  \ , \qq
\d_\Sigma P_{\a\b}= \d_\Sigma X^{\a\b}=\d_\Sigma \t^\a =0\ , \la{ss}
\ee
where the parameter is antisymmetric: $\Sigma_{\a\b}=-\Sigma_{\b\a}$ and
$\e_0$ is defined in \eq{ecs}. The action \eq{na} has also fermionic
symmetries. Firstly, it is invariant under the global supersymmetry
transformations
\be
\d_\e \t^\a = \e^\a\ , \qq \d_\e X^{\a\b} = \e^{(\a} \t^{\b)}\ , \qq
\d_\e P_{\a\b} = 0\ ,\qq \d_\e e_{\a\b} = 0\ , 
\ee
which clearly realize the algebra \eq{alg}. The action \eq{na} also has
local $\k$-symmetry given by
\bea
&& 
\d_\k \t^\a = P^{\a\b}\, \k_\b\ ,\nn\\
&&
\d_\k X^{\a\b} = \t^{(\a}\, \d_\k \t^{\b)}\ , \nn\\
&&
\d_\k e_{\a\b}  = 2 ( \k_\b\,d\t_\a + \e_0\,\k_\a\,d\t_\b)\ , \nn\\
&&
\d_\k P_{\a\b} = 0\ . \la{nk}
\eea
Recall that $ C^T=\e_0\,C$. These transformations close on-shell on the
bosonic $\L$-transformations and $\Sigma$-transformations:
\be
[\d_{\k_{(1)}},\d_{\k_{(2)}}]= \d_\L + \d_\Sigma \ , \la{closen}
\ee
where the composite gauge transformation parameters are
\bea
&&
\L_{\a\b} = 
\left( \k^{(2)}_\b~P_\a{}^\c~\k^{(1)}_\c
  + \e_0\,\k^{(2)}_\a~P_\b{}^\c~\k^{(1)}_\c \right)-(1\lra 2)\ , \nn\\
&&
\Sigma_{\a\b} = 4\e_0\,\k^{(1)}_{[\a}~d \k^{(2)}_{\b]}-(1\lra 2) \ .
\eea
We need the $e_{\a\b}$ equation of motion in showing the closure on
$X^{\a\b}$, and vice versa.

The field equations that follow from the action \eq{na} are
\bea
&&
P_{\a\b}\,P^{\b\c}=0\ ,\la{e1}\\
&&
dP_{\a\b}=0\ , \qq  P_{\a\b}\ d\t^\b =0\ ,  \la{e23}\\
&& 
dX^{\a\b}-\t^{(\a} d\t^{\b)} +e^{\c(\a}\, P^{\b)}{}_\c =0 \ . \la{e4}
\la{neom}
\eea
In particular \eq{e1} implies 
\be
\det~\, P=0\ ,
\ee
which is the familiar BPS condition. We will come back to this point
below.

The last equation can be solved for $P$ yielding the result
\be
P= -2\e_0 e^{-1} \left(E\,\wedge\,\Pi\right)\ , \la{solp}
\ee
where
\be
E \equiv \left({1\over 1+ e^{-1}}\right)\ ,
\ee
and the definition
$$
e^n  \wedge \Pi\,\equiv\,e^n\,\Pi\,e^{-n}\ ,
$$
is to be applied to every term that results from the expansion of $E$ in
$e$. Substituting \eq{solp} into the action \eq{na}, we find 
\be
I =-2 \int d\tau\ \tr\, \left[e^{-1}\left(E\wedge\Pi \right)\right]^2\ .
\ee

Going back to the first order form of the action \eq{na}, it is possible
to introduce a constant mass parameter $m$ by shifting everywhere the
momenta $P_{\a\b}$ occur (including the $\k$-symmetry transformations)
as\\

\noindent{$Massive\ Model: \qq P_{\a\b}\ 
\ra\ P_{\a\b}+ m\,C_{\a\b}$.\\

The charge conjugation matrix $C_{\a\b}$ has to be symmetric, i.e.
$\e_0=1$, for this to make sense.
 
\bigskip


\subsection{ A Multi-Superparticle Reduction of the Model }


There exists an interesting reduction of \eq{na} yielding an
action analogous to that of Sec. 2.  Consider the ansatze 
\be
P= \C\ ,\qq X_i^\m = -\tr~ X V_i^\m\ . \la{newdefs}
\ee
where the spinor indices are suppressed. Recall the definitions
\eq{defc} and \eq{defv} of $\C $ and $ V_i^\m $, respectively. The
definitions \eq{newdefs} lead to the action
\be
I=\int d\tau \left( P^i_\m dX_i^\m -\tb \C d\t 
	+\ft12 e\,(\det~M) \right) \ , \la{rna}
\ee
where we have defined $e\equiv (-1)^{n(n-1)/2}\, C^{\a\b}\,e_{\a\b}$, dropped a
total derivative term $(n-1) d(\tr\, X\C)$ and used the lemma $\C^2=
(-1)^{n(n-1)/2}\,{\rm det}\,M $. The $\L$-symmetry transformations
\eq{nl} reduce to
\be
\d_\L e = d \L\ , \qq \d_\L X_i^\m = -\L~Cof~(M)_{ij}~P^{j\m}\ ,
\qq \d_\L P^i_\m=0\ , \qq \d_\L \t =0\ .
\ee
The $\Sigma$-symmetry \eq{ss} becomes trivial, while the $\k$-symmetry
transformations \eq{nk} take the form
\be
\d_\k \t = \C\,\k\ , \qq
\d_\k X_i^\m = \tb\,V_i^\m\,(\d_\k \t)\ ,\qq
\d_\k e = -4(-1)^{n(n-1)/2}\,d \tb\,\k\ , \qq
\d_\k P^i_\m = 0\ . \la{rnk}
\ee

The action \eq{rna} and its symmetries are similar to those of the model
discussed in the previous section, but they are not quite the same. The
last term in the new action is different and it gives the on-shell
condition
\be
\det~M =0\ , \la{dm}
\ee
as opposed to $M=0$ of the previous section. Recall that $M^{ij} \equiv
P^i_\m P^{j\m}$. 

For two particles, for example, the condition \eq{dm} means that either
${\vec P_1} = \l {\vec P_2}$, where $\l$ is an arbitrary constant, or
${\vec P_i} \cdot {\vec P_j}=0\ (i,j=1,2)$. In the first case $M$ has
rank one and $P^{\m\n}=0$, while in the second case $M$ has rank zero
and $P^{\m\n} \ne 0$. To satisfy \eq{dm} in general, either two or more
momenta should be parallel, in which case $P^{\m_1\cdots \m_n}=0$, or
they should all be perpendicular to each other in which case $M=0$ and
$P^{\m_1\cdots \m_n} \ne 0$. 

\bigskip


\subsection{An Example in $(2,2)$ Dimensions}


It is useful to consider a simple example to see what the action \eq{na}
and the condition \eq{e1} mean. Therefore, let us consider a $(2,2)$
dimensional spacetime with pseudo Majorana spinors. Then, we have the
expansion
\be
P=\c^\m P_\m + \c^{\m\n} P_{\m\n}\ .
\ee
Curiously, there are ten coordinates in total. Substituting this into
the action \eq{na}, and denoting fields that occur in the $\c$-matrix
expansion of the antisymmetric $e_{\a\b}$ by $(e,\phi,\phi_\m)$, we
obtain
\be
I=\int d\tau \left[ P_\m \Pi^\m + P_{\m\n} \Pi^{\m\n}
+e\left(P^\m P_\m -2P^{\m\n} P_{\m\n}\right) 
+ \e^{\m\n\rho\sigma} \left(\phi P_{\m\n} + \phi_\m
P_\n\right)\,P_{\rho\sigma} \right]\ ,
\ee
where
\be
\Pi^\m=dX^\m-\tb\c^\m d\t\ ,\qq \Pi^{\mu\nu}=dX^{\m\n}-\tb\c^{\m\n} d\t \ .
\ee
Note, in particular, the $\k$-symmetry transformation of $\t$:
\be
\d_\k \t =(\c^\m P_\m + \c^{\m\n} P_{\m\n})\,\k\ .
\ee 
The constraints on momenta resulting from the $(e,\phi,\phi_\m)$
equations of motion, equivalent to the single equation $(P^2)^{\a\b}=0$,
are
\be
P^\m P_\m -2P^{\m\n} P_{\m\n}=0\ , \qq 
P_{[\m}\,P_{\r\s]} = 0\ , \qq
P_{[\m\n}\,P_{\r\s]} = 0\ . \la{c123}
\ee
There are a number of ways to satisy these conditions. For example,
introducing a vector $Q_\m$, we have the solution
\be
P_{\m\n}= P_{[\m}\,Q_{\n]}\ ,
\qq 
(P\cdot Q)^2= P^2 (Q^2-1)\ .
\ee
Observe that the vectors $P$ and $Q$ do not have to be null or
orthogonal, though they can be so as a special case. Another solution is
obtained by using two mutually orthogonal null vectors $P^i_\m\,(i=1,2)$
as follows
\be
P_\m=0\ , \qq P_{\m\n} = \e_{ij}\ P^i_\m P^j_\n\ ,
\qq
P^i_\m\,P^{j\m}=0 \ ,
\ee
corresponding to the two-particle model of Sec. 2.

\bigskip


\section{Conclusions}


We have presented a simplified form of the gauge and $\k$-symmetry
transformations of an $n$ superparticle system in $(d-n,n)$ dimensions.
One can consider a version of the model in which there are $n$ fermionic
variables and all the variables have multi-time dependence. In that
case, one can replace $\theta$ by $(\theta_1+\cdots +\theta_n)/n$ and
$\p$ by $(\p_1+\cdots +\p_n)/n$. If one then takes the variables $X_i$
and $\theta_i$ to depend only on $\tau_i$, for example
$\theta_1(\tau_1)$, $X^2(\tau_2)$, etc, then one otains the model
constructed in \cite{rs1}.

In Sec. 3, we have generalized the model discussed in Sec. 2 to a rather
universal one which realizes all the brane charges of the superalgebra
and gives rise to an on-shell condition that includes the BPS condition.
In this model, one need not set the $n$\,th rank brane charge into a
product of $n$-momenta, though this is a partricular solution to the
on-shell condition.

The meaning of physical degrees of freedom and the nature of target
space wave equations in presence of the $p$-form coordinates requires a
better understanding of the representation theory behind the general
superalgebra \eq{alg}. Since the model is supersymmetric, whatever the
degrees of freedom are, they must form a representation of the
underlying superalgebra that survives the BPS condition. We refer the
reader to \cite{b2} for a discussion of how these kind of superalgebras
may be used in unifying the perturbative and nonperturbative states of
$M$-theory.

The string and higher brane generalization of the new model should be of
great interest. In fact, an attempt was made sometime ago to introduce
$p$-form coordinates in trying to build a new type of string action (in
eleven dimensions) \cite{tc}, but the problem of how to achieve
$\k$-symmetry was never resolved \cite{es1}. If one takes the
superparticle limit \cite{bps} of $D=11$ supermembrane, then
one obtains the equations for a massless superparticle, which differ
from the new model presented in Sec. 3.1. The $D=10$ twistor
superparticle model of \cite{sol} is more similar to our model. In
either case, the open problem is how to construct a superbrane action in
the ``maximally $p$-form extended'' superspace, in such a way that it
will give the modfel presented here in the particle limit.

The new model presented here should also give some clues for the
realization of the $M$-algebra super $p$-form charges \cite{bs1,bs2,es2}.
The role of the new coordinates in the realization of duality
symmetries, and as a separate development, in a covariant matrix
formulation of $M$-theory, are some of the other aspects of the new
model that merit further study.

\bigskip


\noindent{\large\bf Acknowledgements}


\bigskip

We are indebted to I. Bars, C.S. Chu, R. Percacci and D. Sorokin for
helpful discussions. One of the authors (E.S.) thanks the Abdus Salam 
International Center for Theoretical Physics for hospitality. 

\pagebreak


\noindent{\Large \bf Appendix A}

\bigskip

\noindent{\large\bf Reduction of the $\kappa$-Symmetry and Trivial Gauge
Transformations}


\bigskip

In order to facilitate comparison of the symmetries presented in Sec. 2,
and those of \cite{deli2}, we take a particular form of the $\k_i$
symmetry given by
\be
\k_i= \ft{1}{n+9} \c_\m V_i^\m  \k \ ,
\ee
which defines the single $\k$-symmetry parameter $\k$. The
transformations \eq{rk} now take the form
\bea
\d_\k \t &=& n\C~\k \ ,\nn\\
\d_\k X_i^\m &=& -\tb V_i^\m (\d_\k \t)\ , \nn\\
\d_\k P_i^\m &=& 0\ , \nn\\
\d_\k e_{ij} &=& -4~Cof~(M)_{ij}~\kb~d\t\ ,
\la{bk}
\eea
where we have used the lemmas
\bea
&&
(\c^\m P^i_\m)\,(\c_\n V_i^\n) = n(n+9)~\C\ , \nn\\[+0.25cm]
&&
\c_\m V_{(i}^\m V_{j)}{}^\n \c_\n  
= -(-1)^{n(n+1)/2}~(n+9)^2~Cof~(M)_{ij}\ ,\nn\\[+0.25cm]
&&
Cof~(M)_{ij} \equiv \ft{1}{(n-1)!}\,\e_{ii_2\cdots i_n} 
\e_{jj_2\cdots j_n}\,M^{i_2 j_2}\cdots M^{i_n j_n}\ . 
\eea
We can use the field equations in the last transformation rule so that
$\d_\k e_{ij}=0$, but we shall not do so in order to compare our results
with those of \cite{deli2}. We will show that the bosonic gauge
symmetries \eq{bn} and the $\k$-symmetry transformations \eq{bk} are
equivalent to those given in \cite{deli2}. To do this, we shall make use
of trivial gauge transformations that are proportional to equations of
motion.

In any field theory involving a collection of fields $\Phi^A$, there
always exist a trivial gauge symmetry that is proportional to the
equations of motion:
\be
\d_\w \Phi^A = \w^{AB} {\d {\cal L}\over \d \Phi_B}\ ,
\ee
where the arbitrary and possibly field dependedent transformations
parameters are only required to be graded antisymmetric:
$\w^{AB}=-(-1)^{AB}\,\w^{BA}$. In our case, we have the set of fields
\be
\Phi^A \equiv (e_{ij}, X_i^\m, P^i_\m, \t^\a)\ .
\ee
The field equatios we will need for the purposes of this section are
\bea
&& 
\d {\cal L} /\d X_i^\m  = - dP^i_\m \equiv R^i_\m \ , \nn\\
&&
\d {\cal L} /\d P_i^\m  = 
	dX_i^\m +\tb V_i^\m d\t-e_{ij} P^{j\m} \equiv S_i^\m  \ , \nn\\
&&
\d {\cal L}/ \d \t  = 2 \C~ d\t + (d\C)~\t \equiv \psi  \ .\la{e}
\eea
Now, consider a special subset of these transformations, involving the
$(X,P)$ equations of motion, given by
\be
\d_\w X_i^\m = \w_i{}^j~S_i^\m\ , \qq
\d_\w P^i_\m = -R_\m^i~\w_i{}^j\ . \la{w1}
\ee
Choosing the transformation parameter $\w_i{}^j$ as
\be
\w_i{}^j= \L_{ik} e^{kj}\ , 
\la{param1}
\ee
where $e^{ij} \equiv (e^{-1})^{ij}$, we find that the sum of the
$\L$-transormations \eq{bn} and the $\w$-transformation \eq{w1} with
parameter \eq{param1} give 
\be
\d e_{ij} = d\L_{ij}\ , \qquad
\d X_i^\m = \L_{ik}~e^{kj}~(dX_j^\m +\tb V_j^\m \t) \ ,\qquad
\d P^i_\m = e^{ik}\L_{kj}~dP^j_\m\ , \qquad 
\d\t = 0\ . \la{bgt}
\ee
These are precisely the bosonic gauge transformations of \cite{deli2}. 
Note that while the  $\L$-transformations \eq{bn} close off-shell, the 
combined $(\d_\L + \d_\w)$ transformations \eq{bgt} close on-shell.

Next we turn to the $\k$-transformation rules \eq{rk}. It will be
sufficient to consider the $\w$-transformations that involve the
$(X,P,\t)$ equations of motion 
\be
\d_\w \t = \w^{i\m}~S_{i\m}\ , 
\qq
\d_\w X_{i\m} = \w_{i\m}{}^{j\n}~S_{j\n}\ ,
\qq
\d_\w P^{i\m} = {\bar\psi}~\w^{i\m}-R^{j\n}~\w_{j\n}{}^{i\m}\ , 
\la{w2}
\ee
where we have suppressed the spinor indices on $\w^{i\m}$. Calculating
the sum of the $\k$-transformations \eq{bk} and the $\w$-transformations
\eq{w2} with parameters
\be
\w^{i\m} = e^{ij} V_j^\m~\k\ , 
\qq
\w_{i\m}{}^{j\n} = -\tb~V_{i\m} e^{jk} V_k^\n~\k\ , \la{param2}
\ee
we find that the combined $\d_\k + \d_\w$ transformations yield 
\bea
\d \t &=& e^{ij} (dX_i^\m +\tb V_i^\m \t) V_{j\m} \k \ ,
\nn\\
\d  X_i^\m &=& -\tb V_i^\m (\d_\k \t)\ , 
\nn\\
\d P_i^\m &=& 2~d\tb~\C e^{ij} V_{j\m} \k\ ,
\nn\\
\d_\k e_{ij} &=& -4~Cof~(M)_{ij}~\kb~d\t \ .
\la{bbk}
\eea
These are precisely the $\k$-transformations of \cite{deli2}, apart from
a sign factor in certain dimensions in the last equation (see 
Appendix B for the relevant symmetry properties of the $\c$-matrices ). 

\pagebreak


\noindent{\Large \bf  Appendix B}

\bigskip

\noindent{\large\bf Properties of Spinors and $\c$-Matrices in Arbitrary
Dimensions}


\bigskip

Here we collect the properties of spinors and Dirac $\c$-matrices in
$(s,n)$ dimensions where $s(n)$ are the number of space(time)
coordinates. The Clifford algebra is $\{\c_\m,\c_\n\}=2 \eta_{\m\n}$,
where $\eta_{\m\n}$ has the signature in which the time-like directions
are negative and the spacelike directions positive. The possible reality
conditions on spinors are listed in Table 1, where $M, PM, SM, PSM$ stand for
Majorana, pseudo Majorana, symplectic majorana and pseudo symplectic
Majorana, respectively \cite{kt,ss}
\footnote
{Corrections to formulae in p. 5\,\&\,6 of \cite{ss}: 
(a) Change eq. (2) to: $(\C^{d+1})^2=(-1)^{(s-t)(s-t-1)/2}$ (our $n$ here is 
denoted by $t$ in \cite{ss}), 
(b) eq. (3) should read $\C^\dagger_\m=(-1)^t A\C_\m A^{-1}$, 
(c) change the last formula in eq. (5) to $B=CA$, 
(d) multiply eq. (6) with $\e$ on the right hand side, 
(e) the prefactor in eq. (8) should be $2^{-\left[d/2\right]}$, 
(f) interchange the indices $\n_1$ and $\n_2$ in the last term of eq. (9). 
Note that here we have let $ C\ra C^{-1}$ relative to \cite{ss}. }
(see below). An additional chirality condition can be imposed for
$s-n=0\ mod\ 4$. 

The symmetry properties of the charge conjugation matrix $C$ and $(\c^\m
C)_{\a\b}$ are listed in Table 2. The sign factors $\e_0$ and $\e_1$
arise in the relations
\be
C^T=\e_0 C\ , 
\qq  
(\c^\m C)^T = \e_1 (\c^\m C)\ .
\la{sp}
\ee
This information is sufficient to deduce the symmetry of 
$(\c^{\m_1\cdots \m_p} C)_{\a\b}$ for any $p$, since the symmetry
property alternates for $p~{\rm mod}~2$. In any dimension with $n$
times, one finds
\be
(\c^{\m_1\cdots \m_p} C)^T= \e_p\,(\c^{\m_1\cdots \m_p} C)\ , 
\qq\quad
\e_p \equiv \e\,\eta^{p+n}\,(-1)^{(p-n)(p-n-1)/2}\ ,
\ee
where $\eta$ is a sign factor. Note that $\e_n=\e$ and
$\e_{n-1}=-\e\eta$. All possible values of $(n,s,p)$ in which $\e_p=+1$,
i.e. the values of $p$ for which $\c^{\m_1\cdots \m_p} C$ is symmetric
(the antisymmetric ones occur for $mod\ 4$ complements $p$) are listed
in Table 3. Other useful formulae are:
\be
\c_\m^T = (-1)^n\,\eta~C^{-1} \c_\m C\ , 
\qq
\c_\m^*= \eta B\c_\m B^{-1}\ ,
\qq
\c_\m^\dagger = (-1)^n A C_\m A^{-1}\ .
\ee
We can choose $A=\c_0\c_1\cdots \c_{n-1}$. Note that $A=BC$.
The chirality matrix in even dimensions $d$ can be defined as
\be
\c_{d+1}\equiv\pm (-1)^{(s-n)(s-n-1)/4}\, \c_0\c_1\cdots \c_{d-1}\ ,
\qq (\c_{d+1})^2=1\ .
\ee
We can choose the sign in the first equation such that the overall factor in
front is $+1$ or $+i$. Using the matrix $B$, we can express the reality
conditions on a (pseudo) Majorana spinor $\psi^* = B\psi$ and a
(pseudo)symplectic Majorona spinor as $(\psi_i)^* = \Omega^{ij}B\psi_j$,
where $\Omega^{ij}$ is a constant antisymmetric matrix satisfying
$\Omega_{ij}\Omega^{jk}=-\d_i^k$ and the index $i$ labels a pseudo-real
representation of a given Lie algebra which admits such a representation
(e.g. the fundamental representations of $Sp(n)$ and $E_7$).
\\

\begin{tabular}{lr}
\begin{minipage}[t]{4in}
\begin{tabular}{|c|c|c|c|}
\hline
 {}  &    {}  &      {}      &       {}       \\
$\e$ & $\eta$ & $(s-n)\ mod\ 8$ &   $Spinor\ Type $  \\
\hline\hline
 $+$   &    $+$   &   0,1,2      &       M       \\
\hline
 $+$   &    $-$   &   6,7,8      &       PM      \\
\hline
 $-$   &    $+$  &    4,5,6      &       SM      \\
\hline
 $-$   &    $-$  &    2,3,4      &       PSM      \\
\hline
\end{tabular}
\vspace{0.5cm}

\indent{\bf Table 1:\ }{\it Spinor types in $(s,n)$ Dimensions}
\end{minipage}
\end{tabular}


\vspace{0.5in}


\begin{tabular}{lr}
\begin{minipage}[t]{4in}
\begin{tabular}{|c|c|c|}
\hline
{} & {}  & {} \\
$n\ mod\ 4$ & $\e_0$ & $\e_1$ \\
\hline\hline
0   & $+\e$      &   $+\e\eta$      \\
\hline
1   & $-\e\eta$  &   $+\e$          \\
\hline
 2   & $-\e$     &   $-\e\eta$      \\
\hline
 3   & $+\e\eta$  &  $-\e$         \\
\hline
\end{tabular}
\vspace{0.5cm}

\indent{\bf Table 2:\ }{\it Symmetries of 
$C$ and $(\c^\m C)$ in $(s,n)$ dimensions.} 
\end{minipage}
\end{tabular}

\vspace{0.5in}


\begin{tabular}{l}
\begin{minipage}[t]{3.5in}
\begin{tabular}{||c|c|c||c|c|c||}
\hline\hline
 {}  &    {}  &      {}      &        
 {}  &    {}  &      {}     \\
$n\ mod\ 4$ & $ s\ mod\ 8 $ & $p\ mod\ 4$   & 
$n\ mod\ 4$ & $ s\ mod\ 8 $ & $p\ mod\ 4$    \\
\hline\hline
${\bf 1}$ &  ${\bf 1},2,3$  &  ${\bf 1},2$ & 
${\bf 3}$ &  ${\bf 3},4,5$  &  ${\bf 3},4$  \\
\hline
    & ${\bf 1},7,8$   &  ${\bf 1},4$ &     
    & ${\bf 3},1,2$   &  ${\bf 3},2$  \\    
\hline
    & ${\bf 5},6,7$   &  ${\bf 3},4$ &     
    & ${\bf 7},8,1$   &  ${\bf 1},2$  \\   
\hline
    & ${\bf 5},3,4$   &  ${\bf 3},2$ &     
    & ${\bf 7},5,6$   &  ${\bf 1},4$  \\  
\hline\hline
${\bf 2}$ & ${\bf 2},3,4$ & ${\bf 2},3$ & 
${\bf 4}$ & ${\bf 4},5,6$ & ${\bf 4},1$  \\
\hline
    & ${\bf 2},8,1$   &  ${\bf 2},1$ &  
    & ${\bf 4},2,3$   &  ${\bf 4},3$  \\
\hline
    & ${\bf 6},7,8$   &  ${\bf 4},1$  &
    & ${\bf 8},1,2$   &  ${\bf 2},3$  \\
\hline
    & ${\bf 6},4,5$   &  ${\bf 4},3$  &
    & ${\bf 8},6,7$   &  ${\bf 2},1$  \\
\hline\hline
\end{tabular}
\vspace{0.5cm}

\indent{\bf Table 3:\ } {\it Symmetric $(\c^{\m_1\cdots \m_p}C)$
in $(s,n)$ dimensions.}\\
{\it In bold cases, an extra Weyl condition is possible. }\\
\end{minipage}
\end{tabular}

\pagebreak


\end{document}